\documentclass[aps,prb,twocolumn]{revtex4}
\usepackage{epsf}
\usepackage[intlimits]{amsmath}
\usepackage{amssymb}
\usepackage{graphicx}

\newcommand{\tcc}{TlCuCl$_3$}

\begin{document}

\title {
A mean-field theory for
strongly disordered non-frustrated antiferromagnets
}

\author{Heidrun Weber}
\author{Matthias Vojta}
\affiliation{\mbox{Institut f\"ur Theorie der Kondensierten Materie,
Universit\"at Karlsruhe, 76128 Karlsruhe, Germany}}
\affiliation{Institut f\"ur Theoretische Physik,
Universit\"at K\"oln, 50937 K\"oln, Germany}
\date{Sep 25, 2006}

\begin{abstract}
Motivated by impurity-induced magnetic ordering phenomena in spin-gap materials like \tcc,
we develop a mean-field theory for strongly disordered antiferromagnets,
designed to capture the broad distribution of coupling constants
in the effective model for the impurity degrees of freedom.
Based on our results,
we argue that in the presence of random magnetic couplings the conventional
first-order spin-flop transition of an anisotropic antiferromagnet
is split into two transitions at low temperatures,
associated with separate order parameters along and perpendicular
to the field axis.
We demonstrate the existence of either a bicritcal point or a critical endpoint
in the temperature--field phase diagram,
with the consequence that signatures of
the spin flop are more pronounced at elevated temperature.
\end{abstract}

\maketitle


\section{Introduction}

The interplay of magnetism and disorder is a fascinating field of
research in condensed matter physics.
Various non-trivial low-temperature phases, like spin glasses, Bose glasses,
random singlet phases etc., and associated phase transitions
have been studied in both theory and experiment.
A particularly interesting manifestation of quantum effects
is impurity-induced magnetism in quantum paramagnets:
The starting point is a Mott insulator with a finite spin gap
which separates elementary spin excitations from the spin singlet ground
state; examples are
SrCu$_2$O$_3$, CuGeO$_3$, PbNi$_2$V$_2$O$_8$, SrCu$_2$(BO$_3$)$_2$, KCuCl$_3$,
or \tcc.
As demonstrated in recent experiments,\cite{exp1,exp2,exp3,exp3b,exp4,exp5,sitetlcucl,dopeins}
{\em non-magnetic} impurities, replacing magnetic ions in such quantum paramagnets,
can induce effective magnetic moments.
These impurity-induced moments reveal themselves in a Curie-like behavior of the
uniform susceptibility, $\chi \propto C/T$, at intermediate temperatures.
Remarkably, in the presence of three-dimensional couplings
these induced moments can order at sufficiently low temperatures,
thus changing the spin-gapped paramagnetic ground state of the pure compound
into an magnetically long-range ordered state upon doping.

On the theoretical side, the appearance of effective moments
upon doping vacancies into the spin system is well understood in principle.
It occurs in systems with confined spinons, i.e., elementary $S=1$ excitations;
for spin $1/2$ systems it is best visualized in terms of broken singlet bonds
where one spin is replaced by a vacancy.
The liberated spin $1/2$ is confined to the vacancy
at low energies, resulting in an effective spin $1/2$ moment.\cite{sigrist,icmp}
(In contrast, in host systems with elementary $S=1/2$ excitations,
i.e., deconfined spinons, no moments are generated by introducing vacancies.)
This theoretical picture has been supported by various numerical
studies, in particular on spin chain\cite{bruce} and ladder
systems.\cite{elbio,Miyazaki,Yasuda}

A number of theoretical works also addressed impurity-induced
antiferromagnetic ordering.
On the one hand, numerical simulations on finite-size systems of spin gap magnets
containing impurities studied signatures of magnetic ordering.\cite{Yasuda,wessel,rosc1,rosc2}
On the other hand, analytical approaches\cite{fabrizio,melin,rsrg1,rsrg2}
typically start from an effective spin-$1/2$
model involving the impurity-induced moments $\vec S_i$ only:
\begin{equation}\label{heff}
\mathcal{H}_{\rm eff} = \sum_{ij} J_{ij} \vec S_i \cdot \vec S_j,
\end{equation}
where the interaction $J_{ij}$ between two impurity moments depends
on their distance $r_{ij}$ as $J_{ij} \propto \exp(-r_{ij}/\xi)$, where $\xi$ is the magnetic
correlation length of the host material.
Due to the random locations of the impurities the system shows a broad
distribution of coupling values $J_{ij}$.
Furthermore, on bipartite lattices the sign of $J_{ij}$ will alternate
as function of the Manhattan distance between $i$ and $j$.
This implies that classically all bonds can be satisfied with a Ne\'el-type
arrangement of the effective moments.
In other words, Eq.~(\ref{heff}) defines a strongly disordered non-frustrated
quantum magnet.
Real-space renormalization group studies\cite{rsrg1,rsrg2} indicate that the ground state
of the model (\ref{heff}) with quantum spins 1/2 shows long-range
order for any concentration of impurity moments;
this is supported by numerical simulations of vacancy-doped quantum paramagnets
with confined spinons, which display magnetic order up to the percolation
threshold.\cite{Yasuda,wessel,rosc1,rosc2}
Thus, although the systems under consideration are long-range-ordered antiferromagnets,
their properties can be expected to be strongly different from that of antiferromagnets
without quenched disorder, due to the broadly distributed $J_{ij}$.

The purpose of this paper is to introduce a generalized mean-field theory,
which takes into account the broad distribution of coupling constants in
Eq. (\ref{heff}).
The central idea is to parameterize the spins according to their coupling
strength to the environment, i.e., by their sum of coupling constants to all other
spins, $\bar J$, see Eq.~(\ref{jbar}) below.
For each $\bar J$ a separate mean-field parameter will be introduced,
leading to integral equations replacing the standard self-consistency
relations.
Although such a mean-field theory misses certain aspects of disorder
physics, like localization phenomena, we will demonstrate that it captures
various distinct properties of magnets described by Eq.~(\ref{heff}).
For instance, the overall behavior of the order parameter as function of
temperature is significantly different from non-disordered magnets and from
conventional mean-field theory.
In particular, we discuss the physics in an external field the presence
of a magnetic anisotropy, relevant for most real materials.
Here, a spin-flop transition is expected to occur for fields parallel
to the easy axis, which indeed has been observed in TlCu$_{1-x}$Mg$_x$Cl$_3$.\cite{sitetlcucl}
We argue that strong disorder leads to an interesting temperature evolution
of the spin-flop physics: at low temperature the transition is generically
split into two (with at least one of them being continuous), whereas
at elevated temperature a single first-order transition is restored.

The bulk of the paper is organized as follows:
In Sec.~\ref{sec:mf} we describe our generalized mean-field theory,
together with the numerical procedure to solve the mean-field equations.
Sec.~\ref{sec:phd} discusses the symmetries and possible phases of the
model in the presence of a field parallel to the easy axis, and
presents temperature--field phase diagrams obtained from the mean-field theory.
In Sec.~\ref{sec:pt} we take a closer look at the phase transitions,
in particular at the spin-flop transition.
A comparison to experiments and available numerical results for vacancy-doped magnets
is given in Sec.~\ref{sec:comp}.
A brief outlook concludes the paper.


\section{Mean-field theory}
\label{sec:mf}

\subsection{Parameterization}

In standard mean-field theory, the many-body problem is reduced
to one single-spin problem in an effective field.
Following this idea in the presence of disorder requires to consider
distinct effective fields for all spins.
Physicswise, we expect spins to behave differently if they have
different couplings to their environment; spins in a similar
environment may behave in a similar fashion.
This is the basis for the main simplification of our mean-field
theory:
We will parameterize the spins by the coupling sum $\bar J$, defined as
\begin{equation}\label{jbar}
\bar J_i = \sum_j J_{ij} (-1)^{i-j+1} = \sum_j |J_{ij}|.
\end{equation}
Thus we replace the spin variables $\vec S_i$ by $\vec S(\bar J)$.
The factor $(-1)^{i-j+1}$ accounts for the sublattice structure of the bipartite
lattice, with $(i-j)$ being the Manhattan distance,
and the second identity follows from the alternating sign of the
coupling $J_{ij}$.
With this definition, $\bar J_i$ is the magnitude of the effective field
on spin $i$ in a perfectly ordered antiferromagnetic state.

For a description of the magnetic interactions in terms of the coupling sum,
we introduce a probability distribution $P(\bar J)$ according to
\begin{equation}\label{pj}
P(\bar J) = \frac 1 N \sum_i \delta({\bar J} - \bar J_i)
\end{equation}
where $N$ is the number of spins.
Further, we need an interaction function $f(\bar J_1,\bar J_2)$,
defined as
\begin{equation}\label{fjj}
f(\bar J_1,\bar J_2) = \frac 1 {N P(\bar J_1)P(\bar J_2)}
\sum_{ij} |J_{ij}| \delta({\bar J_1} - \bar J_i) \delta({\bar J_2} - \bar J_j) \,.
\end{equation}
which is symmetric w.r.t. to $\bar J_1 \leftrightarrow \bar J_2$ and
fulfills normalization conditions
\begin{equation}
\label{fnorm}
\int d\bar J_2 P(\bar J_2) f(\bar J_1, \bar J_2)=\bar J_1 \,.
\end{equation}

With these definitions, a ferromagnetic Heisenberg model,
$\mathcal{H}=-\sum_{ij} J_{ij} \vec S_i \cdot \vec S_j$, takes the form
\begin{equation}
\mathcal{H} = -N \int d \bar J_1 d \bar J_2
P(\bar J_1) P(\bar J_2) f(\bar J_1,\bar J_2) \vec S(\bar J_1) \cdot \vec S(\bar J_2).
\end{equation}
Defining an effective field
\begin{equation}
\vec m(\bar J) = \int d \bar J_2
P(\bar J_2) f(\bar J,\bar J_2) \langle \vec S(\bar J_2)\rangle,
\end{equation}
the mean-field Hamiltonian reads
\begin{equation}
\mathcal{H}_{\rm mf} =
-N \int d \bar J P(\bar J) \left[\vec m(\bar J) + \vec B\right] \cdot \vec S(\bar J),
\label{hmf_fm}
\end{equation}
where we have included an external field $\vec B$.

For an antiferromagnet on a bipartite lattice different mean fields are required for
the two sublattices A and B, which can become inequivalent in the presence of
symmetry breaking and a finite field.
Using $P_A(\bar J)=P_B(\bar J)=P(\bar J)/2$, the effective field becomes
\begin{equation}
\label{mfdef}
\vec m_{A,B}(\bar J)=\pm \int d\bar J_2 P(\bar J_2) f(\bar J,\bar J_2)
\frac{\langle \vec S_A(\bar J_2)\rangle-\langle \vec S_B(\bar J_2)\rangle}{2}.
\end{equation}
The corresponding mean-field Hamiltonian reads
\begin{equation}
\begin{split}
\mathcal{H}_{\rm mf} = -\frac{N}{2}
\int d \bar J P(\bar J&) \left(  \left[ \vec{m}_A(\bar J)+\vec B\right] \vec S_A(\bar J)+\right.\\
&\left. + \left[\vec{m}_B(\bar J)+\vec B\right] \vec S_B(\bar J) \right)
 \end{split}
\label{hmf_afm}
\end{equation}
replacing (\ref{hmf_fm}).

In general, two angles are needed to specify the orientation of a spin.
In the following, we will exclusively consider situations where $\mathcal{H}$
has a U(1) symmetry of rotations about the $z$ axis (see Sec.~\ref{sec:phd}).
Then we can describe the orientation of a spin just by one angle, $\varphi$,
between spin and $z$ axis;
the angle in the $xy$ plane can be fixed to zero, positive (negative) values of
$\varphi$ on the A (B) sublattice account for the staggered in-plane order.
Hence, the expectation value of a spin can be written in the following way:
\begin{equation}
\label{angledef}
\langle \vec S_{A,B}(\bar J)\rangle = \left|\langle \vec S_{A,B}(\bar J)\rangle\right| \left(
\begin{array}{c}
\sin \varphi_{A,B}(\bar{J})\\
0 \\
\cos \varphi_{A,B}(\bar{J})\\
\end{array}
\right).
\end{equation}

The single-spin problem of Eq.~(\ref{hmf_fm}) can be readily solved,
yielding the mean-field equations:
\begin{subequations}
\label{mfeq}
\begin{equation}
\cos \varphi_{A,B}(\bar J) = \vec{e_z} \cdot
\frac{[\vec{B}+\vec m_{A,B}(\bar{J})]}{|\vec{B}+\vec m_{A,B}(\bar{J})|} \,,
\end{equation}
and
\begin{equation}
\left|\langle \vec{S}_{A,B} (\bar{J}) \rangle \right| =
s \tanh \frac{\left|\vec{B}+\vec m_{A,B}(\bar{J})\right|}{k_B T/s} \,.
\end{equation}
\end{subequations}
The amplitude equation (\ref{mfeq}b) has been written for a spin with two
states $\pm s$, appropriate for quantum spins $s=1/2$ -- this will be
used in the numerical calculations.
For continuous classical spins the $\tanh()$ needs to be replaced by a Brillouin
function as usual.

In the case of $P(\bar J)=\delta(\bar J -\bar J_0)$ and $f(\bar J_1,\bar J_2)=\bar J_0$
the mean-field equations (\ref{mfdef},\ref{mfeq}) reduce to the self-consistency equation of the
familiar Weiss mean-field theory,
with $\bar J_0=zJ$ for a Hamiltonian with a nearest-neighbor coupling strength $J$ and a
coordination number of $z$.

\subsection{Magnetic anisotropy}

In this paper we consider the formally simplest source of magnetic anisotropy,
namely an anisotropic exchange interaction of easy-axis type
(the behavior in the presence of
a Dshyaloshinski-Moriya interaction is expected to be similar).
Thus, in the Hamiltonian we perform the replacement
\begin{equation}\label{haniso}
J_{ij} \vec S_i \cdot \vec S_j
~\rightarrow~
J_{ij} \left[\beta(S_i^x S_j^x + S_i^y S_j^y) + S_i^z S_j^z \right]
\end{equation}
with an anisotropy constant $\beta<1$;
we keep the coupling to the external field as $\vec B \cdot \vec S$.
We note that in the context of impurity-induced magnetism, e.g., in \tcc\
some complications arise (which we will ignore here):
The anisotropy of the effective interaction will depend on the external
field and the interaction itself; furthermore the form of the
field coupling will be modified, leading to an anisotropic $g$ tensor.

The anisotropy according to Eq.~(\ref{haniso}) requires the following replacement
in the mean-field equations (\ref{mfeq}):
\begin{equation}
\vec m_{A,B}
~\rightarrow~
\left(
\begin{array}{c}
\beta m^x_{A,B}\\
\beta m^y_{A,B} \\
m^z_{A,B}\\
\end{array}
\right)
\end{equation}
where $m^{xy}$ and $m^z$ denote the components perpendicular and parallel
to the easy axis.

\subsection{Choice of input parameters}

The described mean-field theory requires the coupling distribution $P(\bar J)$,
Eq.~(\ref{pj}), and the interaction function $f(\bar J_1,\bar J_2)$, Eq.~(\ref{fjj}),
as input.
Both functions are given by the underlying microscopic model.
We have numerically determined $P(\bar J)$ and $f(\bar J_1,\bar J_2)$
for an effective model of impurity-induced order in quantum paramagnets, Eq.~(\ref{heff}),
as described in the Appendix.

For the actual mean-field calculations we found it more convenient to
use plausible model (i.e. fitting) functions instead.
However, a difficulty arises here: $P(\bar J)$ and $f(\bar J_1,\bar J_2)$
cannot be chosen independently, because the normalization conditions
(\ref{fnorm}) cannot be easily fulfilled while preserving the symmetry w.r.t.
$\bar J_1 \leftrightarrow \bar J_2$.
We have therefore resorted to the following construction.
From an arbitrary symmetric ``generating'' function $g(\bar J_1,\bar J_2)$
we define the functions $P(\bar J)$ and $f(\bar J_1, \bar J_2)$ according to:
\begin{equation}
\label{Pg}
P(\bar{J})=\frac{a}{\bar{J}}\int d\bar J_2 g(\bar J, \bar J_2)
\end{equation}
and
\begin{equation}
f(\bar{J}_1,\bar J_2)=a\frac{ g(\bar{J}_1,\bar J_2)}{P(\bar{J}_1)P(\bar J_2)} \,.
\end{equation}
Choosing the normalization factor as
\begin{equation}
\label{eq:adef}
a^{-1} =
\int d\bar{J}_1 \frac{1}{\bar{J}_1}
\int d\bar J_2  g(\bar{J}_1,\bar J_2),
\end{equation}
all normalization conditions are fulfilled.

\begin{figure}[t]
\resizebox{220pt}{!}{
\includegraphics{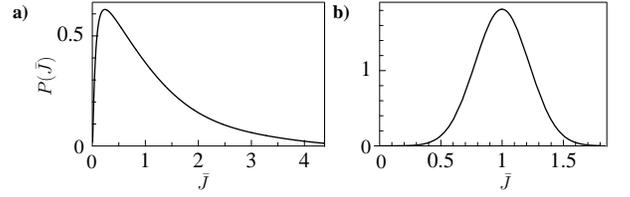}
}
\caption{
Model functions for the coupling-sum distribution $P(\bar J)$
obtained from different generating functions $g(\bar J_1,\bar J_2)$:
a) Lognormal-like distribution from Eq.~(\ref{logng}),
with the parameters $\sigma_2=0.699$, $\mu_L=-0.171$, $\sigma_L=1.3$, and
$\bar J_{\text{max}}=4.317$.
b) Gauss-like distribution from Eq.~(\ref{gaussg}),
with the parameters $\mu_1=0.59$, $\sigma_1=0.289$, $\sigma_2=0.193$, and
$\bar J_{\text{max}}=1.929$.
In both cases, $\int d\bar J P(\bar J)=1$.
}
\label{fig:konP}
\end{figure}

As detailed in the Appendix, in vacancy-doped magnets the function $P(\bar J)$
will be peaked at a value which increases with impurity concentration,
and it will be increasingly asymmetric at low concentrations, with
a tail to higher $\bar J$.
Thus, among others, we have employed the following generating functions
$g(\bar J_1, \bar J_2)$.
\begin{equation}
\label{logng}
\begin{split}
g(\bar{J}_1,\bar J_2)&=
\exp\left( -\frac{(\ln(\bar{J}_1)-\mu_L)^2}{2 \sigma_L^2}\right)
\exp\left( -\frac{(\ln(\bar J_2)-\mu_L)^2}{2 \sigma_L^2}\right)\\
&\times \exp\left( -\frac{(\bar{J}_1-\bar J_2)^2}{2 \sigma_2^2}\right)
\theta(\bar J_{\text{max}}-\bar{J}_1) \theta(\bar J_{\text{max}}-\bar J_2 )
\end{split}
\end{equation}
generates a Lognormal-like distribution $P(\bar J)$ which captures the coupling distribution
for a small impurity concentration.
In contrast,
\begin{equation}
\label{gaussg}
\begin{split}
g(\bar{J}_1,\bar J_2)&=
\bar{J}_1 \bar J_2
\exp\left( -\frac{(\bar{J}_1-\mu_1)^2}{2 \sigma_1^2}\right)
\exp\left( -\frac{(\bar J_2-\mu_1)^2}{2 \sigma_1^2}\right)\\
&\times\exp\left( -\frac{(\bar{J}_1-\bar J_2)^2}{2 \sigma_2^2}\right)
\theta(\bar J_{\text{max}}-\bar{J}_1) \theta(\bar J_{\text{max}}-\bar J_2 )
\end{split}
\end{equation}
leads to a Gauss-like distribution $P(\bar J)$, corresponding to higher impurity concentrations.
The $\mu_i$ and $\sigma_i$ are free parameters.
Examples for the resulting $P(\bar J)$ are shown in Fig. \ref{fig:konP};
here and in the following the $\bar J$ values are scaled such that
the mean value of $\bar J$ is unity.
In Fig. \ref{fig:numP} below we show $P(\bar J)$ results of the numerical simulations for
comparison.

To model an easy-axis situation, we have chosen the anisotropy parameter $\beta = 0.9$.


\section{Phase diagrams}
\label{sec:phd}

\subsection{Symmetries and phases}

We will restrict our attention to magnetic fields parallel to the easy axis
in which case interesting spin-flop physics arises.
Then, the symmetry of the antiferromagnetic Hamiltonian (\ref{hmf_afm})
in the presence of a field is U(1) $\times$ Z$_2$ (whereas for $B=0$ we have
U(1) $\times$ Z$_2$ $\times$ Z$_2$ which becomes SU(2) $\times$ Z$_2$ in the absence
of exchange anisotropy).
Here we have assumed a unit cell size of 2 sites, and
the Z$_2$ symmetry corresponds to the exchange of the two sublattices.

The possible phases are easily enumerated:
(i) An Ising phase, where spins on the A (B) sublattice point preferrentially up (down).
This breaks the Z$_2$ symmetry, but leaves the U(1) rotations about the $z$ axis intact.
(ii) A canted phase, where the spins order spontaneously perpendicular to the field, and
have a finite component in the field direction (equal for both sublattices) as well.
Z$_2$ and U(1) are broken, but a combination of sublattice exchange and 180$^{\rm o}$ rotation
is an intact Z$_2$ symmetry.
(iii) A mixed phase where Z$_2$ and U(1) are fully broken.
(iv) A disordered phase with no symmetry breaking. For non-zero field the spins
point in the field direction only.

\begin{figure}[t]
\centering\includegraphics[width=3.4in]{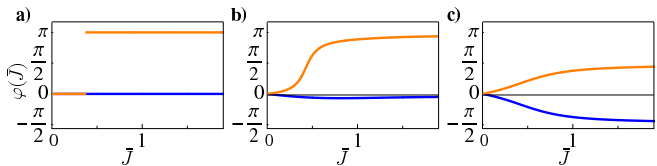}
\caption{
(color online)
Angle configurations $\varphi(\bar J)$ occurring in the ordered phases of the
antiferromagnet with field parallel to the easy axis, with blue/red (dark/gray)
showing $\varphi_A$/$\varphi_B$.
a) Ising phase, b) mixed phase, c) canted phase.
The evolution from a) to c) represents the behavior
at low temperatures upon increasing the field,
see Figs.~\ref{fig:pd1},\ref{fig:pd2} below.
}
\label{fig:phi}
\end{figure}

The order parameters are the components of the staggered magnetization parallel and
perpendicular to the direction of the applied magnetic field, $M_{\text{stagg},z}$ and
$M_{\text{stagg},xy}$.
Further, the ordered phases (i)--(iii) can be nicely characterized by the angles $\varphi$,
Eq.~(\ref{angledef}).
Without quenched disorder, i.e., for $P(\bar J)=\delta(\bar J-\bar J_0)$, we have
(i) Ising: $\varphi_A=0$, $\varphi_B=\pi$;
(ii) canted: $0<\varphi_B=-\varphi_A<\pi$;
(iii) mixed: $\varphi_A<0$, $\varphi_B>0$, $|\varphi_A|\neq|\varphi_B|$.
In the disordered case with broad $P(\bar J)$, the angles become functions of $\bar J$,
with examples for $\varphi_{A,B}(\bar J)$ shown in Fig.~\ref{fig:phi}.

\subsection{Numerical iteration of the mean-field equations}

For given functions $P(\bar J)$ and $f(\bar J_1, \bar J_2)$ and fixed values of
applied magnetic field $\vec B$, temperature $T$, and anisotropy constant
$\beta$ one can iterate the mean-field equations (\ref{mfdef},\ref{mfeq}), using
a linear discretization for the $\bar J$ values.
The initial distributions for the angles $\varphi_{A,B}(\bar J)$ and amplitudes
$|\langle \vec{S}_{A,B} \rangle|$ could be chosen random in principle;
we found it more convenient to employ distributions corresponding to perfect
Ising or XY order instead.
The mean fields are calculated from Eq.~(\ref{mfdef}); new amplitudes are obtained
from Eq.~(\ref{mfeq}b).
Some care is required with the angle equation (\ref{mfeq}a) in the case of an Ising
initial configuration:
in addition to Eq.~(\ref{mfeq}a) we used an update scheme where
spins on the sublattice B flipping from $\varphi_B=\pi$ to $\varphi_B=0$
are set to $\varphi_B=\pi/2$ by hand in order to mix the symmetry breakings.

These different initial conditions and different update schemes lead to potentially
different fixed-point distributions after convergence is reached. These correspond to
different local minima in the free-energy landscape; comparison of the free energies
then yields the stable phase.

\subsection{Phase diagrams from mean-field theory}

\begin{figure}[tb]
\centering\includegraphics[width=3.2in]{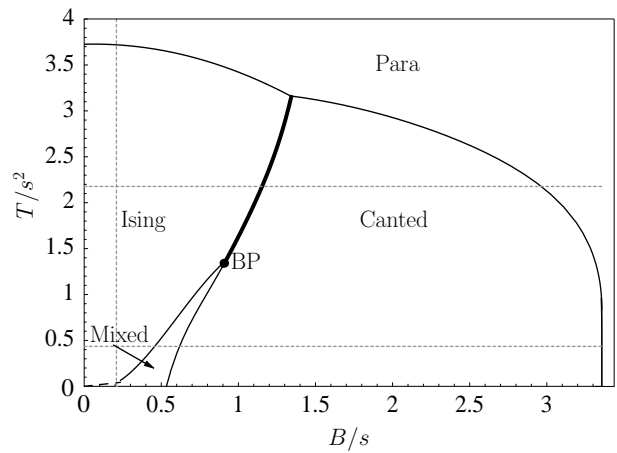}
\caption{
Temperature--field phase diagram of a disordered easy-axis antiferromagnet,
obtained from the present mean-field theory,
for the Lognormal-like distribution of coupling constants $P(\bar J)$ as
shown in Fig.~\ref{fig:konP}a.
The calculation was done for spins with eigenvalues $\pm s$, hence the axes
are scaled by appropriate powers of $s$.
First-order transitions (thick lines) were determined from crossing points of
the free energy, whereas second-order transitions were obtained from extrapolations
of the order parameters. BP is a bi-critical point.
Magnetization data along the dotted gray lines will be shown
in Figs.~\ref{fig:mag1}, \ref{fig:magstagg} below.
}
\label{fig:pd1}
\end{figure}

\begin{figure}[tb]
\centering\includegraphics[width=3.2in]{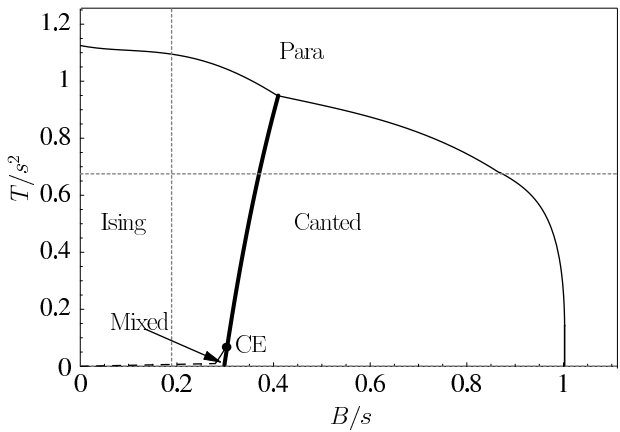}
\caption{
Temperature--field phase diagram as in Fig.~\ref{fig:pd1},
but for the Gauss-like distribution $P(\bar J)$ shown in Fig.~\ref{fig:konP}b.
(CE is a critical endpoint.)
}
\label{fig:pd2}
\end{figure}

In Figs. \ref{fig:pd1}, \ref{fig:pd2} we show representative phase diagrams
for the disordered easy-axis antiferromagnet with a longitudinal field,
obtained from solving the mean-field equations (\ref{mfdef},\ref{mfeq}).
At zero field, the Ising order, present at low temperatures $T$,
is destroyed at a continuous transition to a paramagnetic phase.
Applying a field to the Ising phase drives various transitions,
resulting in a canted phase at intermediate fields and finally a field-polarized
(disordered) phase at large field.
The main difference to the text-book antiferromagnet is the presence of a
mixed phase. The conventional first-order spin-flop transition is split
into two transitions at low $T$:
At some small field, there is a continuous transition from the Ising to a mixed
phase, where an in-plane staggered magnetization $M_{stagg,xy}$ perpendicular to the field
develops.
Only at a larger field, the Ising order measured by $M_{stagg,z}$ is destroyed,
leading to a spin-flop transition into a canted phase, see
Figs.~\ref{fig:mag1}, \ref{fig:mag2}.
The origin of this behavior lies in the broadly distributed couplings:
Already for small fields, weakly coupled spins with $\bar J \ll B$ cannot sustain the
zero-field Ising order and flip in the field direction, while for strongly coupled spins
($\bar J \gg B$) Ising order is favored.
At low $T$, spins with $\bar J \approx B$ reach their lowest-energy state by canting --
this results in an overall mixed phase and can be nicely seen in Fig.~\ref{fig:phi}b.
(However, at elevated temperatures the in-plane mean-field from these spins alone
is not sufficient to establish a mixed phase.)
At the spin-flop transition
(mixed$\rightarrow$canted at low $T$, Ising$\rightarrow$canted at higher $T$),
the spins with the largest couplings loose their Ising order.
The evolution of the angle distributions $\varphi_{A,B}(\bar J)$ with
increasing field at low $T$ is shown in Fig.~\ref{fig:phi}.

Note that the part of the phase transition line between Ising and mixed phases
at small $T$ and $B$ (dashed) is difficult to extract numerically for a linear
discretization of $\bar J$.
However, by a comparison of energies it is easy to prove that at $T=0$, $B>0$ the Ising phase
is always unstable towards the mixed phase for distributions $P(\bar J)$ which
are non-zero for arbitrarily small $\bar J$.

The two phase diagrams in Figs. \ref{fig:pd1}, \ref{fig:pd2} differ in the extensions
of the mixed phase, and in the character of the phase transition line between mixed and
canted phase (see below).
Clearly, the deviations from the textbook spin-flop behavior of non-disordered
antiferromagnets are most pronounced for broad distributions $P(\bar J)$.

Both the N\'eel temperature $T_N$ and the critical field $B_c$ where
long-range order is destroyed, scale with the mean value of the exchange constant
(which is set to unity in our calculations).
Similarly, the flop field $B_{\rm flop}$ scales as $(1-\beta)^{1/2}$ times the mean
exchange.
Furthermore, an asymmetric distribution $P(\bar J)$ as the one in Fig.~\ref{fig:konP}a
leads to a larger $T_N$, $B_c$, $B_{\rm flop}$ compared to a symmetric one with the
same mean $\bar J$,
the reason being that the scales are primarily determined by the spins with
large $\bar J$.


\section{Phase transitions and critical properties}
\label{sec:pt}

\subsection{Magnetizations and critical exponents}

\begin{figure}[t]
\resizebox{220pt}{!}{
\includegraphics{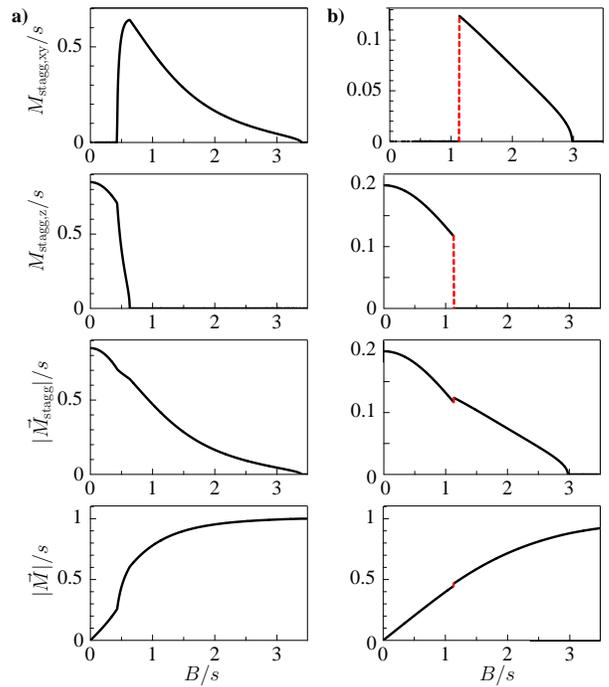}
}
\caption{
(color online)
Magnetization curves for the Lognormal-like distribution $P(\bar J)$, Fig.~\ref{fig:konP}a,
at
a) $T/s^2=0.435$ and
b) $T/s^2=2.177$.
The red (dashed) part indicates the jump at the first-order transition.
}
\label{fig:mag1}
\end{figure}

\begin{figure}[t]
\resizebox{220pt}{!}
{
\includegraphics{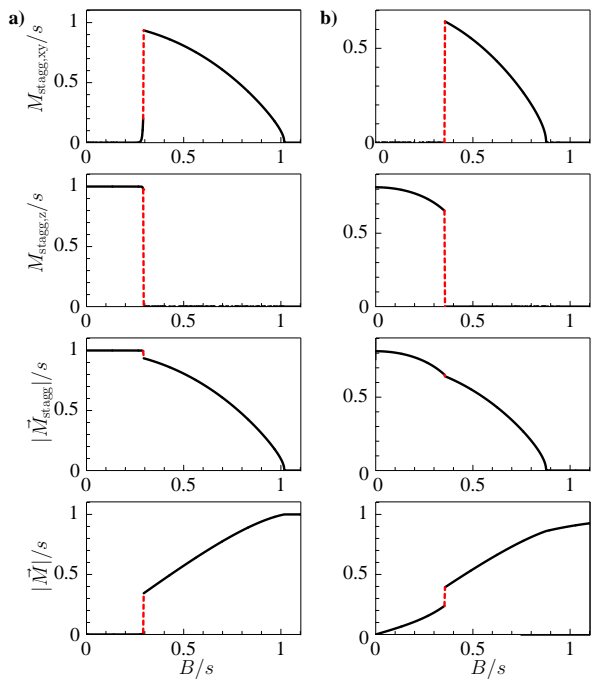}
}
\caption{
(color online)
As Fig.~\ref{fig:mag1}, but for the Gauss-like distribution $P(\bar J)$, Fig.~\ref{fig:konP}b,
at
a) $T/s^2=0$ and
b) $T/s^2=0.675$.
}
\label{fig:mag2}
\end{figure}

The Figs. \ref{fig:mag1}, \ref{fig:mag2} show curves for the staggered
magnetizations ($M_{\text{stagg},xy}$ and $M_{\text{stagg},z}$) as well as the
total magnetization $M$,
along the gray dotted lines in the phase diagrams in Figs. \ref{fig:pd1}, \ref{fig:pd2}.
The role of $M_{\text{stagg},xy}$ and $M_{\text{stagg},z}$ as order parameters
can be clearly seen.

\begin{figure}[b]
\resizebox{220pt}{!}{
\includegraphics{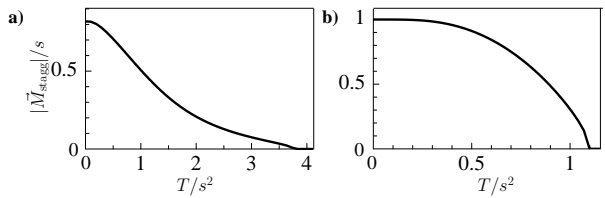}
}
\caption{
Staggered magnetization for
a) the Lognormal-like distribution $P(\bar J)$, Fig.~\ref{fig:konP}a, at $B/s=0.209$, and
b) the Gauss-like distribution $P(\bar J)$, Fig.~\ref{fig:konP}b, at $B/s=0.189$.
}
\label{fig:magstagg}
\end{figure}

Fig. \ref{fig:magstagg} shows the total staggered magnetization vs. temperature
for both coupling-sum distributions.
The transition to the disordered phase at high $T$ (or $B$) is always continuous.
The corresponding order-parameter exponent $\beta$,
defined through $M_{\text{stagg}} \propto (T_c-T)^\beta$,
should be $1/2$ in conventional mean-field theory.
Interestingly, Fig. \ref{fig:magstagg}a shows an overall behavior very different
from this standard square-root law.
This is again due to the broadly distributed couplings: Close to $T_c$ the
magnetization is effectively only carried by a small fraction of the spins with
large $\bar J$ -- note that the distribution $P(\bar J)$ corresponding to
Fig.~\ref{fig:magstagg}a has a pronounced tail at larger $\bar J$.
We note that asymptotically close to the phase transition
standard mean-field behavior is restored within the present approach,
with exponent $\beta=1/2$.

\subsection{Spin-flop transition}

In a non-disordered easy-axis antiferromagnet, a first-order spin-flop transition
from an Ising to a canted phase occurs upon increasing the field,
leading to a jump in the total magnetization.

In the present situation with quenched disorder, a mixed phase with both
$M_{\text{stagg},xy}$ and $M_{\text{stagg},z}$ non-zero occurs at low temperatures.
The spin-flop transition splits; for small disorder (a narrow distribution $P(\bar J)$)
the mixed$\rightarrow$canted transition remains of first order, but becomes second-order
at larger disorder, whereas the Ising$\rightarrow$mixed transition is always continuous.
As the mixed phase only exists at low $T$, the conventional first-order spin-flop
transition is restored at elevated temperatures.
This has the remarkable consequence that the jump in the magnetization is most
pronounced at intermediate $T$, namely at the position of the critical end point
or the bicritical point, respectively (see Figs. \ref{fig:mag1}, \ref{fig:mag2}).
In general, the magnetization jump is larger in situations with less disorder because
the collective spin flop is carried by a larger fraction of spins here.


\section{Relation to vacancy-doped magnets}
\label{sec:comp}

In the previous sections, we have described a general mean-field theory
for disordered non-frustrated antiferromagnets.
We now discuss the applicability to impurity-induced magnetic order
in quantum paramagnets.

\subsection{Experiments}

The low-energy physics of Mg-doped \tcc\ may be expected to be well described
by a Hamiltonian of the form (\ref{heff}), provided that the impurity concentration
is small.
Magnetization measurements\cite{sitetlcucl}
have indicated the presence of a spin-flop transition at 1.8\,K for fields parallel to
the easy axis [2,0,1] at a field of approximately 0.35\,T, significantly below the
field corresponding to the bulk gap, 6\,T.
Interestingly, the spin-flop field seems to be almost independent of
the impurity concentration (in the measured range of 0.8 -- 2.5\%).
Our theory does not offer an easy explanation for that:
As discussed above, the spin-flop field $B_{\rm flop}$ is roughly proportional to the
mean exchange (for fixed anisotropy parameter $\beta$), which would
result in a concentration dependence of $B_{\rm flop}$.
Within the effective model (\ref{heff}), it is hard to envision a mechanism leading to
a concentration-independent flop field, therefore we speculate that correlations
between the impurities beyond this effective model play a role here. We also consider
it possible that $B_{\rm flop}$ will actually decrease for impurity concentrations
smaller than the ones measured.
In this context we note that the experiments of Ref.~\onlinecite{fujisawa}
mapped out the phase diagram for \tcc\ doped with 1\% Mg and
showed the zero-field impurity-induced ordered phase to be continuously connected
to the high-field bulk ordered phase.
Together with recent theoretical studies \cite{mikeska,rosc1}, this suggests
that the impurity and bulk energy scales are not well separated there, i.e.,
the impurity concentration is too high to allow for a description
using the effective model (\ref{heff}).

Further magnetization measurements on Mg-doped \tcc\ would also be interesting
regarding the temperature dependence of the spin-flop physics:
Our theory predicts interesting behavior for very low temperatures,
where the spin-flop transition should split.
This requires measurements down to e.g. 1/10th of the ordering
temperature $T_N$; the experiments of Ref.~\onlinecite{sitetlcucl}
have $T\gtrsim T_N/2$.

\subsection{Numerical results from Quantum Monte Carlo simulations}

Numerical simulations using Quantum Monte Carlo techniques can
go beyond the effective model (\ref{heff}) and study the full system, i.e.,
quantum paramagnet plus vacancies.
Those calculations have been reported in Refs.~\onlinecite{Yasuda,wessel,rosc1,rosc2},
but all were restricted to the spin-isotropic situation.
These simulations mapped out the complete phase diagram, with distinct
low-field and high-field ordered phases.
Among the interesting aspects are the occurrence of Bose glass phases
near the bulk field-ordered phase \cite{rosc1},
and of a quantum disordered phase at intermediate fields where impurity-induced
transverse order is destroyed and the impurity moments appear to form
a random-singlet-like phase \cite{rosc2}.
Clearly, these properties rely on the one hand on the quantum nature of the
impurity-induced spin-$1/2$ moments and on the other hand on localization effects,
both not captured by our mean-field approach.
Quantum Monte Carlo calculations for the spin-anisotropic case studied by us
would be interesting, but may be difficult due to the small energy scales
involved in the spin-flop physics.


\section{Conclusions}

We have proposed a mean-field theory for strongly disordered magnets,
which takes into account the broad distribution of energy scales in the system.
Parameterizing the spins by their sum of coupling constants,
equivalent to the exchange field in a perfectly ordered state,
yields a continuous set of mean-field equations.
We have applied the formalism to a model for impurity-induced order
in spin-gap quantum magnets, and derived detailed phase diagrams as
function of temperature and external field.
We have shown that the conventional first-order spin-flop transition
generically splits at low temperatures, leaving room for a mixed phase
with both transverse and longitudinal order.

We envision our approach of continuous mean fields to be applicable
to a number of interesting problems involving strong disorder,
like magnetic ordering in dilute magnetic semiconductors \cite{xin},
charge ordering in the presence of strong pinning,
or electronic models treated within modifications of dynamical mean-field
theory (DMFT) \cite{sdmft}.


\acknowledgments

We thank T. Roscilde, W. Uschel, X. Wan, and P. W\"olfle for discussions.
This research was supported by the
DFG Center for Functional Nano\-struc\-tures and the
Virtual Quantum Phase Transitions Institute (Karls\-ruhe).


\appendix
\section{Coupling constants for impurity-induced moments in paramagnets}

Our mean-field theory requires the knowledge of the distribution of
the coupling-constant sums, $P(\bar J)$, and the interaction function
$f(\bar J_1, \bar J_2)$.
We have numerically determined these functions from an effective Hamiltonian
for the impurity-induced moments of the form (\ref{heff}).

The effective interaction between two impurity spins ($\vec{s}_{i,j}$),
coupled to two different spins ($\vec{S}_{i,j}$) of a bulk system
according to
\begin{equation}
\mathcal{H}=\mathcal{H}_{\text{bulk}}+
K\left( \vec{S}_i \cdot \vec{s}_i+\vec{S}_j \cdot \vec{s}_j \right)
\end{equation}
can be determined in perturbation theory in $K$.
In lowest order and in the static approximation,
it is given by the $\omega=0$ bulk susceptibility:
\begin{equation}
J_{\text{eff}}^{\alpha}=
K^2 \langle \langle S_i^{\alpha};S_j^{\alpha}\rangle \rangle (\omega=0).
\end{equation}
For vacancy-induced moments, $K$ can be approximated by a bulk exchange coupling.

To be specific, consider a bulk system consisting of dimers on a
$d$-dimensional hypercubic lattice, with intra-dimer (inter-dimer) coupling
$J_\perp$ ($J_\parallel$).
Using a bond-operator formalism in the linearized (harmonic) approximation\cite{kotov},
we find for the effective coupling in the long-distance limit:
\begin{equation}
\label{jeff}
\begin{split}
J_{\text{eff}} = - \frac{K^2}{2 J_{\parallel}} &(-1)^{\zeta} (-1)^r \xi^{(3-d)/2} 2^{(d-1)/2} \pi^{(d+1)/2}\\
&\times  r^{-(d-1)/2} \exp(-r/\xi)
\end{split}
\end{equation}
where $r$ denotes the distance between the impurity spins,
$(-1)^r$ accounts the alternating sign of the interaction, and
$\zeta=0$ (1) for spins $i,j$ on the same (on different) sites of the
dimer pairs.
Note that the asymptotic behavior $r^{-(d-1)/2} \exp(-r/\xi)$ is generic,
whereas the concrete value of $\xi$ and the prefactor in (\ref{jeff})
depend on the level of approximation used; in our case
$\xi^2 = J_\parallel/(J_\perp - 2d J_\parallel)$.
For the purpose of our numerical simulation, we will employ Eq.~(\ref{jeff})
with $K=J_\parallel$ for all distances $r$ -- this will overestimate couplings at small $r$.
The sign of $J_{\text{eff}}$ leads to a
non-frustrated system on a bipartite lattice.

\begin{figure}[t]
\resizebox{220pt}{!}{
\includegraphics{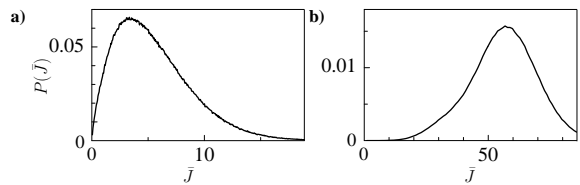}
}
\caption{
Numerical results for the coupling-sum distribution
$P(\bar J)$, in a host magnet with a correlation length of 7.9 lattice spacings,
at impurity concentrations of
a) $0.1\%$ and
b) $1\%$.
(The energy unit for the vertical axis is $J_\parallel$ of the bilayer model;
note that the rather large values of $\bar J$ arise from the inaccurate short-range
behavior of Eq.~(\ref{jeff}).)
}
\label{fig:numP}
\end{figure}

The simulation is done by randomly placing impurities on sites of
a bilayer square lattice (i.e. $d=2$), calculating $P(\bar J)$ and $f(\bar J_1, \bar J_2)$,
Eqs.~(\ref{pj},\ref{fjj}), from the effective coupling constants (\ref{jeff}),
and averaging the result over several impurity configurations.
The bilayer system is assumed to be in its quantum disordered phase,
close to the magnetic ordering transition.
In the linearized bond-operator approach, the phase transition takes place at
$J_{\parallel}/J_{\perp}=1/4$; we choose in the following
$J_{\parallel}/J_{\perp}=0.249$, corresponding to a correlation length of
$\xi\approx 7.9$.
Fig.~\ref{fig:numP} shows the resulting $P(\bar J)$ for two different
impurity concentrations:
In the low-concentration limit, the distribution is strongly asymmetric, with
a tail to high energies.
(Note that in this limit analytical results are available as well, see
e.g. Ref. \onlinecite{mikeska}.)



\begin{thebibliography}{}

\bibitem{exp1} M.~Hase, I.~Terasaki, Y.~Sasago, K.~Uchinokura, and
H.~Obara, Phys. Rev. Lett. {\bf 71}, 4059 (1993).

\bibitem{exp2} S.~B.~Oseroff,
S.-W.~Cheong, B.~Aktas, M.~F.~Hundley, Z.~Fisk, and L.~W.~Rupp, Jr.,
Phys. Rev. Lett. {\bf 74}, 1450
(1995).

\bibitem{exp3} K.~M.~Kojima {\it et al.}, Phys. Rev. Lett.
{\bf 79}, 503 (1997).

\bibitem{exp3b} M. Azuma, Y. Fujishiro, M. Takano, M. Nohara and H. Takagi,
Phys. Rev. B {\bf 55}, R8658 (1997).

\bibitem{exp4} T.~Masuda, A.~Fujioka, Y.~Uchiyama, I.~Tsukada,
K.~Uchinokura, Phys. Rev. Lett. {\bf 80}, 4566 (1998).

\bibitem{exp5} Y.~Uchiyama, Y.~Sasago, I.~Tsukada, K.~Uchinokura, A.~Zheludev,
T.~Hayashi, N.~Miura, and P.~B\"{o}ni, Phys. Rev. Lett.
{\bf 83}, 632 (1999).

\bibitem{sitetlcucl}
A. Oosawa, T. Ono, and H. Tanaka, \prb {\bf 66}, 020405(R) (2002).

\bibitem{dopeins}
A. Oosawa, M. Fujisawa, K. Kakurai, and H. Tanaka,
\prb {\bf 67}, 184424 (2003).

\bibitem{sigrist}
M. Sigrist and A. Furusaki,
J. Phys. Soc. Jpn. {\bf 65}, 2385 (1996).

\bibitem{icmp} S. Sachdev and M. Vojta,
in: Proceedings of the XIII International Congress on Mathematical Physics, London,
eds. A. Fokas {\em et al.}, International Press, Boston (2001).

\bibitem{bruce}
B. Normand and F. Mila, Phys. Rev. B {\bf 65}, 104411 (2002).

\bibitem{Miyazaki} T. Miyazaki, M. Troyer, M. Ogata, K. Ueda, and D. Yoshioka,
J. Phys. Soc. Jpn. {\bf 66}, 2580 (1997).

\bibitem{elbio}
G.~B.~Martins, M.~Laukamp, J.~Riera, and E.~Dagotto, \prl {\bf 78}, 2563 (1997);
M.~Laukamp, G.~B.~Martins, C.~Gazza, A.~L.~Malvezzi, E.~Dagotto,
P.~M. Hansen, A.~C.~Lopez, and J.~Riera, \prb {\bf 57}, 10755 (1998).

\bibitem{Yasuda} C. Yasuda, S. Todo, M. Matsumoto, and H. Takayama,
Phys. Rev. B {\bf 64}, 092405 (2001), and references therein.

\bibitem{wessel}
S.~Wessel, B.~Normand, M.~Sigrist, and S.~Haas, \prl {\bf 86}, 1086 (2001).

\bibitem{rosc1}
T. Roscilde and S. Haas, \prl {\bf 95}, 207206 (2005).

\bibitem{rosc2}
T. Roscilde, cond-mat/0602524 (2006).

\bibitem{fabrizio}
M. Fabrizio and R. Melin, \prl {\bf 78}, 3382 (1997).

\bibitem{melin}
M. Fabrizio, R. Melin, and J. Souletie, Eur. Phys. J. B {\bf 10}, 607 (1999);
R. Melin, Eur. Phys. J. B {\bf 18}, 263 (2000).

\bibitem{rsrg1}
R. Melin, Eur. Phys. J. B {\bf 16}, 261 (2000).

\bibitem{rsrg2}
N. Laflorencie, D. Poilblanc, and M. Sigrist,
\prb {\bf 71}, 212403 (2005);
N. Laflorencie and D. Poilblanc,
J. Phys. Soc. Jpn. Suppl. {\bf 74}, 277 (2005).







\bibitem{fujisawa}
M. Fujisawa, T. Ono, H. Fujiwara, H. Tanaka, V. Sikolenko, M. Meissner, P. Smeibidl,
S. Gerischer, and H. A. Graf,
J. Phys. Soc. Jpn. {\bf 75}, 033702 (2006).

\bibitem{mikeska}
H.-J. Mikeska, A. Ghosh, and A. Kolezhuk, \prl {\bf 93}, 217204 (2004).


\bibitem{xin}
An idea similar to ours has been developed by
X. Wan and R. N. Bhatt (unpublished)
in the context of ordering in dilute magnetic semiconductors.

\bibitem{sdmft}
We note that the so-called statistical DMFT,
V. Dobrosavljevic and G. Kotliar,
\prl {\bf 71}, 3218 (1993),
\prb {\bf 50}, 1430 (1994),
is based on an idea somewhat similar to ours;
however, no parametrization in terms of energy scales is employed,
but instead the full geometry of a finite-size disordered system is kept
(which is mandatory to describe localization phenomena).

\bibitem{kotov} V.~N.~Kotov, O.~P.~Sushkov, Zheng Weihong, and
J.~Oitmaa, \prl {\bf 80}, 5790 (1998).

\end{thebibliography}
\end{document}